\documentclass[12pt,preprint]{aastex}
\usepackage{epstopdf}

\begin{document}

\title{A New Method to Identify Nearby, Young, Low-mass Stars} 

\author{David R. Rodriguez\altaffilmark{1}, M.\ S.\ Bessell\altaffilmark{2}, B.\ Zuckerman\altaffilmark{1}, Joel H.\ Kastner\altaffilmark{3}}

\altaffiltext{1}{Dept.\ of Physics \& Astronomy, University of California, Los Angeles 90095, USA \\
(drodrigu@astro.ucla.edu)}
\altaffiltext{2}{The Australian National University, 
Cotter Road, Weston Creek ACT 2611, Australia}
\altaffiltext{3}{Center for Imaging Science, Rochester Institute of
Technology, 54 Lomb Memorial Drive, Rochester NY 14623}

\begin{abstract}
We describe a new method to identify young, late-type stars 
within $\sim$150~pc of the Earth that employs visual or near-infrared 
data and the GALEX GR4/5 database.
For spectral types later than K5, we demonstrate that the ratio of 
GALEX near-ultraviolet (NUV) to visual and near-IR emission is 
larger for stars with ages between 10 and 100 Myr than for older, main sequence stars.
A search in regions of the sky encompassing the TW~Hya and Scorpius-Centaurus
Associations has returned 54 high-quality candidates for followup.
Spectroscopic observations of 24 of these M1--M5 objects reveal 
Li 6708\AA~absorption in at least 17 systems. 
Because GALEX surveys have covered a significant 
fraction of the sky, this methodology should prove valuable 
for future young star studies.
\end{abstract} 


\section{Introduction}

The last two decades have seen the discovery of a number of 
moving groups and associations near the Earth 
containing young stars with ages in the range $\sim$10
to 100~Myr (for reviews see \citealt{ZS04} and \citealt{Torres:2008}).
By studying stars in these moving groups we can observe 
the evolution of stellar properties as a function of age.
Identifying additional members, particularly fainter, low-mass stars, 
is key to understanding the processes involved 
in the earliest stages of 
stellar, substellar, and massive planet evolution.

Low-mass stars with ages ranging from 10--100~Myr are luminous sources of 
X-rays with respect to their bolometric 
luminosities \citep[and references therein]{Kastner:1997,ZS04,Preibish:2005,SACY}.
Several successful searches for nearby young stars 
exploited their characteristically large X-ray 
activities to identify promising candidates \citep{Kastner:1997,Song:2003,SACY}.
In addition to X-ray activity, however, young low mass stars also 
exhibit larger chromospheric activity than their older counterparts 
(as measured by the $R'_{HK}$ index; 
see \citealt{Mamajek:2008} and references therein).
Higher chromospheric activity is correlated with 
increased ultraviolet emission in stars \citep{Walkowicz:2009}.
Hence, when compared to the older population of low mass stars, young stars will stand out 
as being UV-bright (for example, see Figure~3 in \citealt{Guinan:2009}).

The Galaxy Evolution Explorer satellite (GALEX, \citealt{Martin:2005}) 
has performed a number of large area surveys in the 
far-ultraviolet (FUV, 1344--1786\AA) and near-ultraviolet (NUV, 1771--2831\AA).
The All-sky Imaging Survey (AIS), in particular, has covered $\sim$3/4 of the sky, 
and hence represents an excellent tool to identify UV-bright sources.
However, in order not to damage its detectors, GALEX generally does not 
point within 10 degrees of the galactic plane.
Even with this limitation, GALEX has already explored the regions covered 
by many nearby, young stellar associations \citep{ZS04,Torres:2008}.

A previous study performed in the Taurus and Upper Scorpius regions 
used GALEX to identify candidate young stars \citep{FH10}.
However, these regions are populated by stars with ages $\leq$5~Myr.
We have successfully applied a similar strategy (described in \S~\ref{method}) 
to search for nearby stars with ages 10--100~Myr and with spectral types later than K5
using archival data available in the GALEX GR4/5 data release. 

\section{Method}\label{method}

The search strategy we have employed consists of identifying UV-excess, 
late-type objects using NUV-Optical/NIR colors.
With available proper motion data from the literature, 
we perform a UVW space velocity analysis and the 
best candidate objects are selected for spectroscopic followup.
We describe our method in detail in Sections 2.1--2.3.

\subsection{GALEX (UV-based) Identification of Candidate Nearby, Young Stars} \label{method:uv}

\begin{figure}[htb]
\begin{center}
\includegraphics[width=14cm,angle=0]{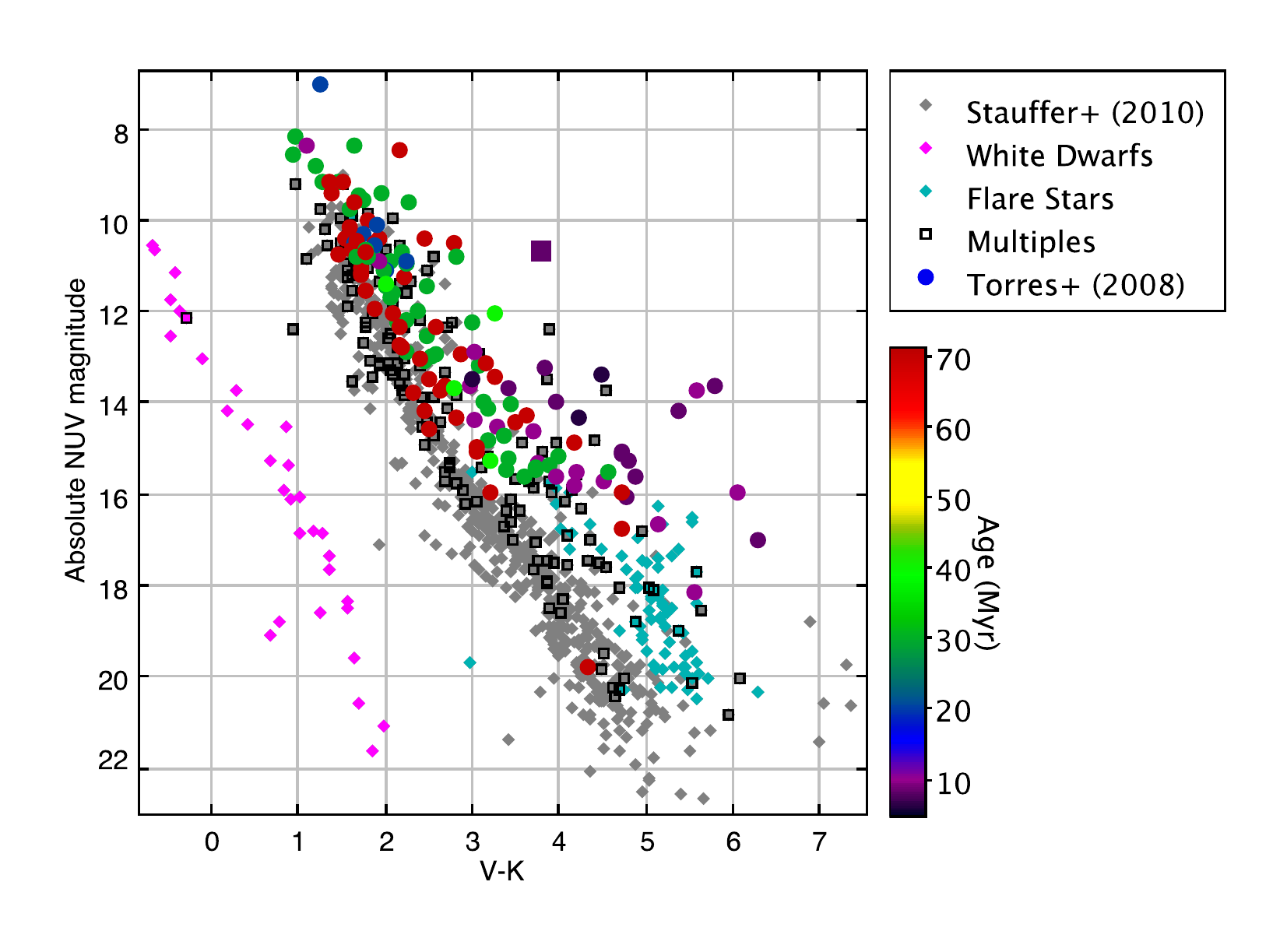}
\end{center}
\caption{
The circles, color-coded by age, are objects in \citet{Torres:2008} that were matched in GALEX.
TW Hya is displayed as a large square.
All other symbols are stars from \citet{Stauffer:2010}, which provides accurate coordinates 
for most of the stars in the Gliese catalog \citep{Gliese:1991}.
Different symbols highlight some of the known multiple, white dwarfs, and flare star systems. 
The grey symbols left of the main sequence are likely to have incorrect distances, 
V-band magnitudes, or both (some such systems were corrected when 
updated information was available).
In general, the Gliese low-mass stars (K5/K7 and later, V--K$\geq3$) 
are older and fainter in NUV than stars 
in young associations by 2 magnitudes on average, 
the notable exception being flare stars and some multiples.
Multiples among the \citet{Torres:2008} stars are not labeled.
}
\label{fig:nuvabs}
\end{figure}

\begin{figure}[htb]
\begin{center}
\includegraphics[width=14cm,angle=0]{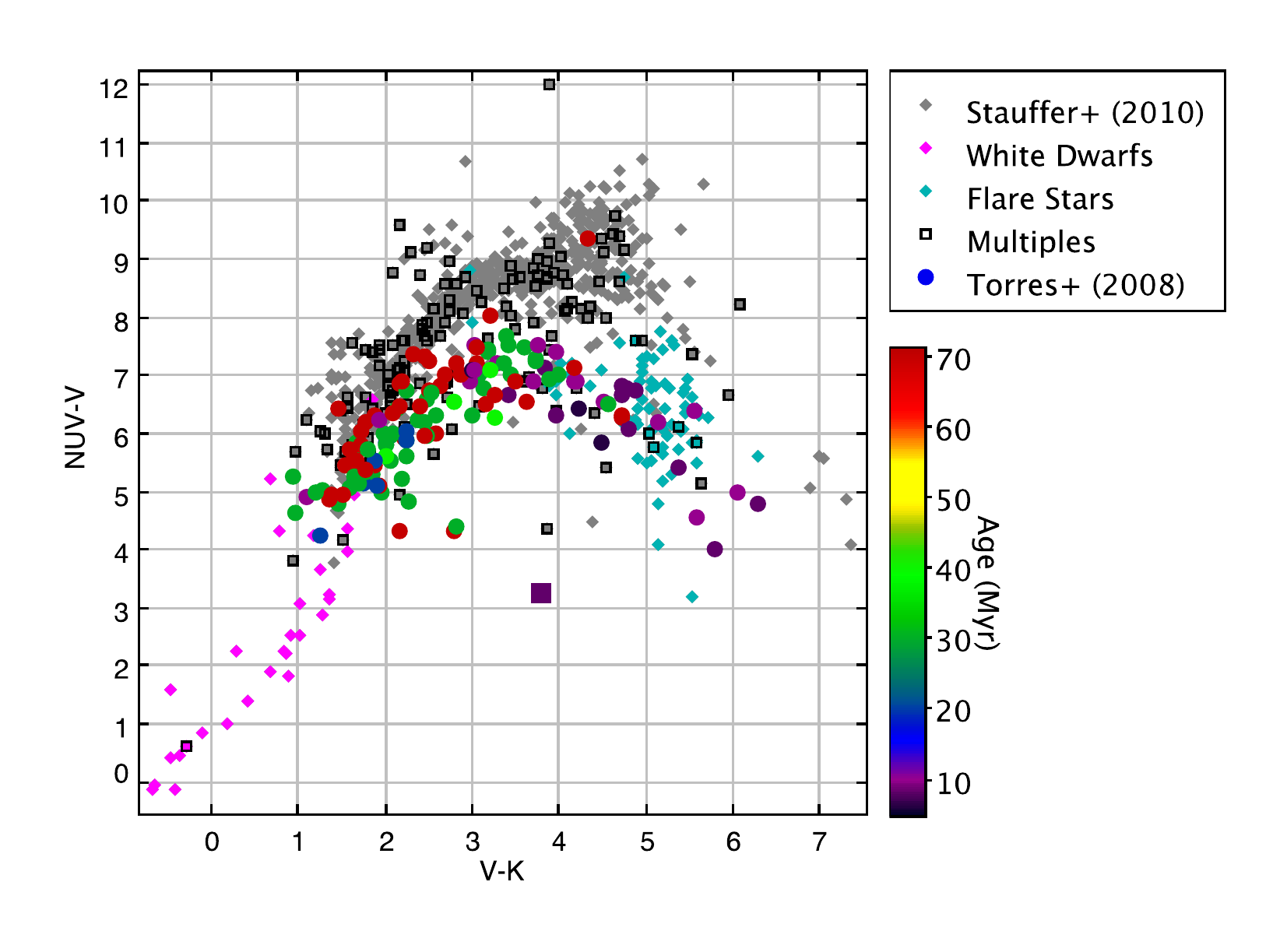}
\end{center}
\caption{
When plotting the members of young associations and the Gliese stars (see Fig.~\ref{fig:nuvabs}) 
in NUV--V color, the distinction between old and young, late-type (K5/K7 and later, V--K$\geq3$) stars become more readily apparent.
Objects with lower NUV--V are brighter in NUV and generally younger.
TW Hya is displayed as a large square.
Putative AB~Dor member BD~+01~2447 is located at V--K$\sim$4.3 and NUV--V$\sim$9.4.
We comment on both these systems in \S~\ref{additional}.
}
\label{fig:nuvv}
\end{figure}

To examine whether 10 to 100~Myr old stars stand out in the ultraviolet, 
we compare the absolute NUV magnitudes for stars in \citet{Torres:2008}, 
which lists members of many of the young, nearby stellar associations, with 
the ``Catalogue of Nearby Stars, 3rd Edition" by \citet{Gliese:1991}; 
most the latter are considerably older than a few 100 Myr.
In particular, we used the catalog provided in \citet{Stauffer:2010}, which 
lists accurate coordinates for most of the stars in \citet{Gliese:1991}.
To identify UV sources we make use of CasJobs\footnote{http://mastweb.stsci.edu/gcasjobs/}, 
an interface to query the GALEX database.
The results are displayed in Figures~\ref{fig:nuvabs} and \ref{fig:nuvv}.
It is clear from these figures that young stars stand out from the older population.
In particular, stars with V--K$\gtrsim$3.5 ($\sim$M0) and ages as old as 70~Myr
(the AB~Dor group in \citealt{Torres:2008}) are brighter in NUV than the older Gliese stars; 
the only exceptions are flare stars (from \citealt{Gershberg:1999} or SIMBAD) 
and some multiple systems (from \citealt{CCDM} and \citealt{SB9}).
GALEX has a PSF FWHM of 5\arcsec~and, hence, cannot resolve binary systems 
with much smaller separations.
Additionally, in some close binary systems (such as RS CVn-type systems) the 
component stars can interact with one another and may produce 
additional UV emission.
The only exception to this trend of brighter NUV magnitudes for young stars is BD~+01~2447, 
classified by \citeauthor{Torres:2008} (\citeyear{Torres:2008}; see also references within) 
as an AB~Dor member, 
with V--K$\sim$4.3, which we discuss later in \S~\ref{additional}.
Fig.~\ref{fig:nuvv}, which displays NUV--V colors for these stars, also shows 
how young stars stand out.
Reddening towards these objects can significantly affect the shorter wavelengths. 
\citet{FH10} apply a reddening curve in which $A_{NUV}/A_V=2.63$; 
however, using the reddening curve of \citet{Cardelli:1989}
and $R_V=3.1$, we obtain $A_{NUV}/A_V=2.96$.
While most Gliese stars are within 25~pc, most of the 
young stars are more distant.
Regardless of the exact value for $A_{NUV}/A_V$, 
extinction towards the younger, more distant objects will  
generally be higher than for the nearby Gliese stars. 
Thus the effect of extinction will be to bring the locus of the young star
populations in Figs.~\ref{fig:nuvabs} and \ref{fig:nuvv} closer to the Gliese old star locus. 
That is, the intrinsic separation in absolute NUV magnitude and in NUV--V
between young and old stars is, if anything, somewhat greater on
average than indicated in Figs.~\ref{fig:nuvabs} and \ref{fig:nuvv}.

Young, early-M stars in these associations (with ages 10--30~Myr) 
tend to have absolute NUV magnitudes between 15--16 (see Fig.~\ref{fig:nuvabs}). 
The 3-$\sigma$ sensitivity for the GALEX AIS survey is 22.0 magnitudes 
for each 100-second exposure, 
which means that these young stars could be detected 
at the 3-$\sigma$ level out to at least 160~pc.
\citet{FH10} estimate that GALEX can detect unextincted photospheres of K5 stars 
out to 140~pc with 300-second exposures, implying absolute NUV magnitudes of about 17.
Mid-K stars with ages of 70~Myr, however, have absolute NUV magnitudes generally brighter 
than 15, suggesting that even at 70~Myr, these stars will be about 2 magnitudes brighter 
than their older counterparts (as illustrated in Fig.~\ref{fig:nuvabs} at V--K$\sim$3).
As previously mentioned, extinction towards these objects will cause them 
to have redder NUV--V colors--- similar to that of older field stars.
Extinction of (at least) $A_V\sim0.7$ magnitudes would be 
required for 70~Myr-old mid-K stars to have 
NUV magnitudes comparable to equally distant, but older, mid-K stars.

\begin{figure}[htb]
\begin{center}
\includegraphics[width=14cm,angle=0]{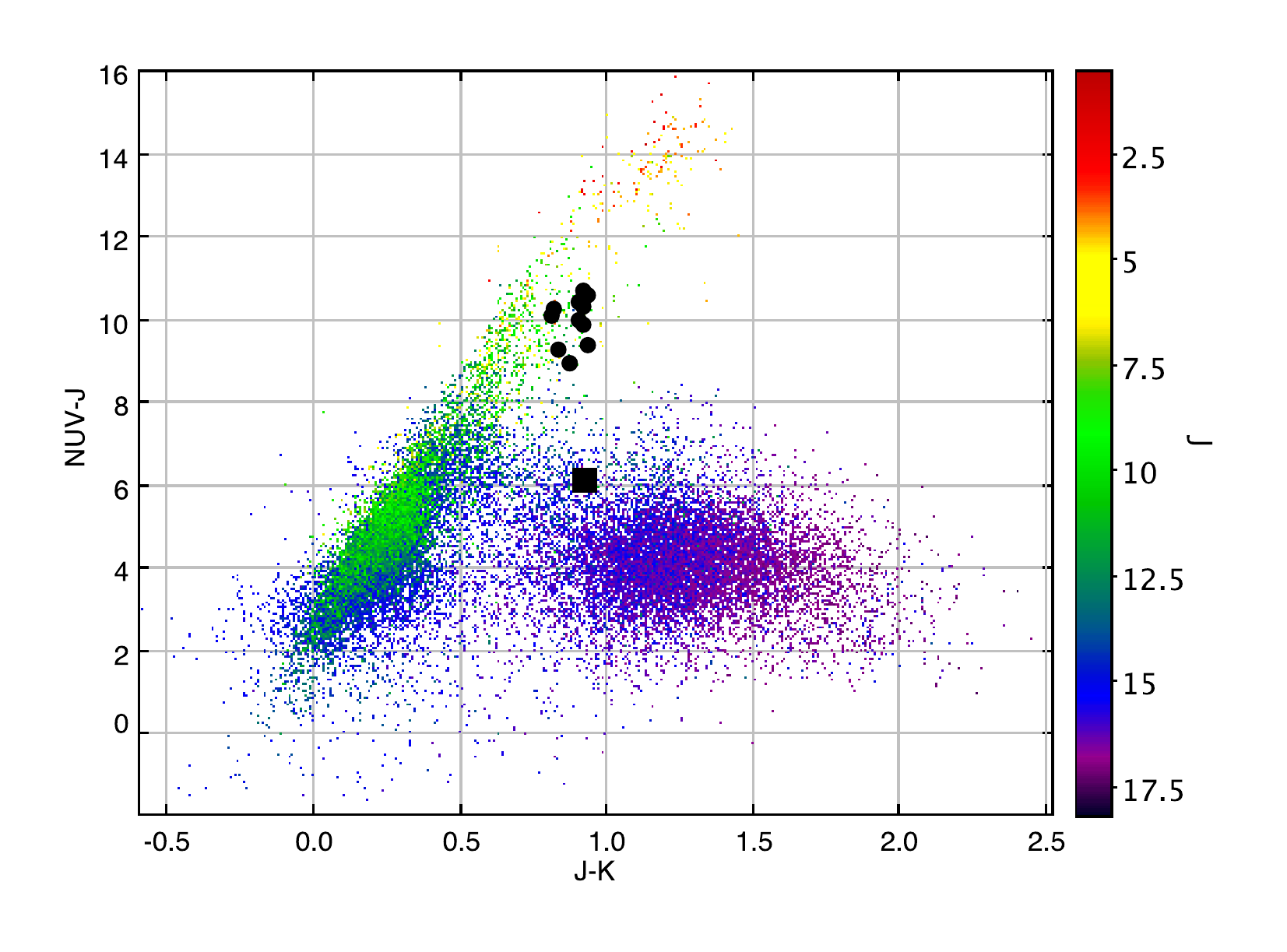}
\end{center}
\caption{Color-color plot for the TWA search.
The colored dots (color based on J magnitude) are the GALEX/2MASS sources, 
while the larger black dots are known TWA members from \citet{Torres:2008}.
The TWA member labeled as a large square is TW Hya itself.
The cloud of sources between about 0.5$<$J--K$<$2 and 2$<$NUV--J$<$7 are generally
distant galaxies and most can be removed from this plot by imposing a cut of J$\leq$14.
All TWA objects lie somewhat below the line of stellar sources 
(ie, they have UV excesses; see Fig.~\ref{fig:ages}).
}
\label{fig:colors}
\end{figure}

\begin{figure}[htb]
\begin{center}
\includegraphics[width=14cm,angle=0]{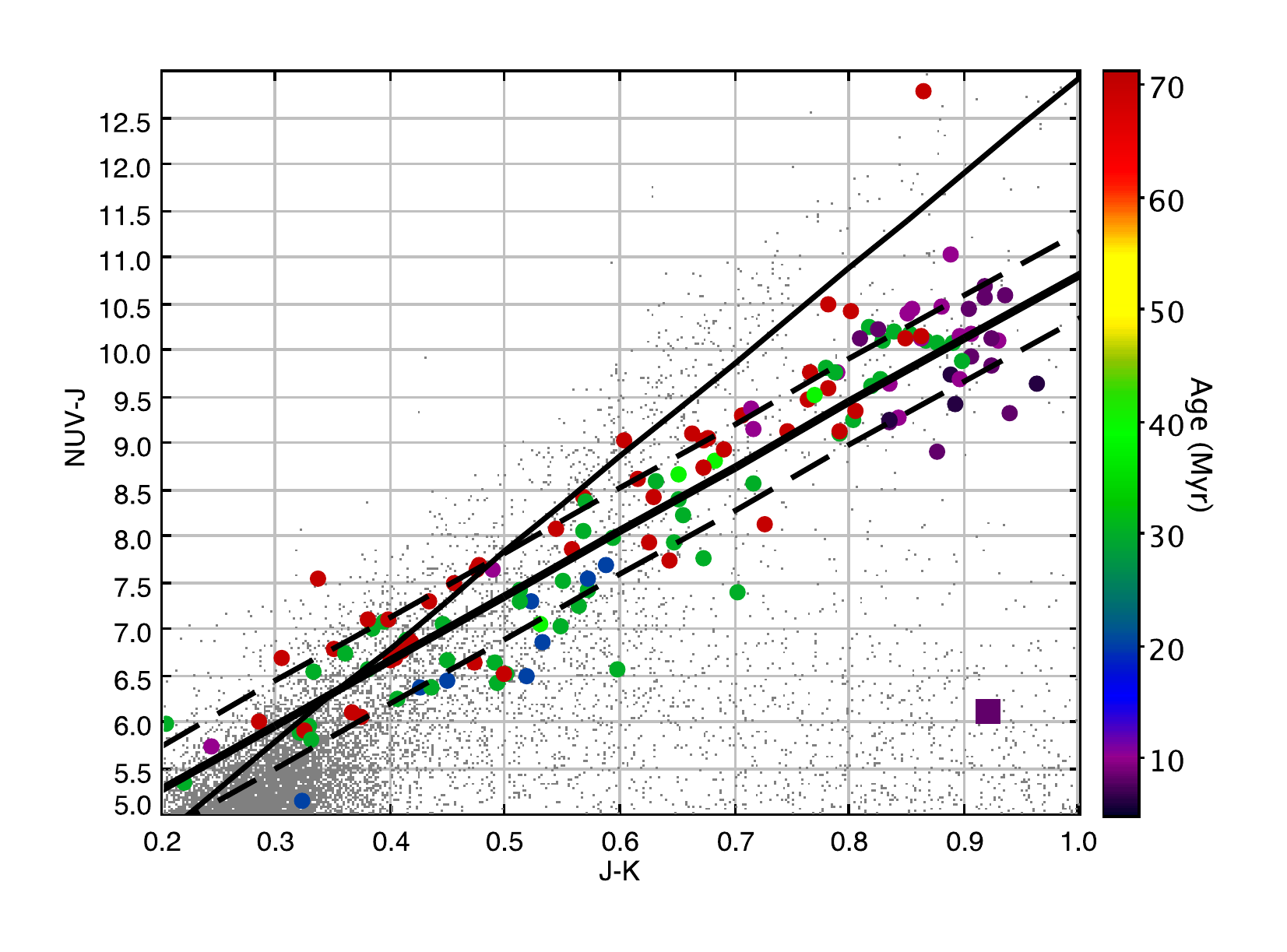}
\end{center}
\caption{
The grey dots are the GALEX/2MASS sources in the vicinity of TWA (see the first line entry to Table~\ref{tab:fields}); those near the bottom right are the upper portion of the galaxy distribution (see Fig~\ref{fig:colors}).
The colored circles are for stars in \citet{Torres:2008}, where the color corresponds
to the age of the association or moving group.
TW Hya is the large purple square.
The thin solid line represents the relations in \citet{FH10} and defines 
the stellar sequence as extrapolated from early-type stars.
The thick solid line is the best-fit line to the young star sample and 
demonstrates (as in Figs~\ref{fig:nuvabs} and \ref{fig:nuvv}) 
how young late-type stars stand out.
Dashed lines outline the 1-$\sigma$ variation among the young stars.
}
\label{fig:ages}
\end{figure}

To see if these 10--100~Myr-old systems also stand out as UV-bright in GALEX-2MASS colors, 
and to carry out a pilot study for the feasibility of a large scale program, we examine 
a region of sky encompassing the TW~Hya Association (TWA).
After defining a suitable search region (see Table~\ref{tab:fields}), 
we select all objects detected in NUV 
(our initial search of the TW Hya Association also required detection in FUV).
These sources are then cross-matched with the 2MASS 
Point Source Catalog \citep{Cutri:2003} with a matching radius of 2\arcsec, 
somewhat larger than the GALEX pointing uncertainty \citep[$\sim$1\arcsec,][]{Morrissey:2007}.
This allows us to produce color-color plots like the one in Figure~\ref{fig:colors}.
In these plots, galaxies tend to lie in a well defined region of space 
(between about 0.5$<$J--K$<$2 and 2$<$NUV--J$<$7) while stars follow a diagonal 
line from J--K$\sim$0 and NUV--J$\sim$2 to J--K$\sim$1 and NUV--J$\sim$13.
The black symbols in Figure~\ref{fig:colors} 
are the TWA stars from \citet{Torres:2008}, 
all of which lie below the stellar sequence; TW Hya is the object with NUV--J$\sim$6.
The stellar locus is well defined by an average of the relations in \citeauthor{FH10} 
(\citeyear{FH10}; see their Equation~2) and we have 
plotted this as a solid line in Figure~\ref{fig:ages}.
Also shown are the 10--100~Myr old stars, 
which can again be distinguished from older field stars in that they 
lie below the stellar sequence (i.e., they have UV excesses).
The thick solid line represents the best linear fit to the locus of young stars, 
and illustrates the difference between the young and old stellar populations, particularly for the lower mass stars (J--K $\geq$ 0.7).
While the separation is not as pronounced as in the NUV--V plots, 
clearly these diagrams should enable one to select candidate 
young, low-mass stars using only GALEX and 2MASS colors.

\begin{figure}[htb]
\begin{center}
\includegraphics[width=14cm,angle=0]{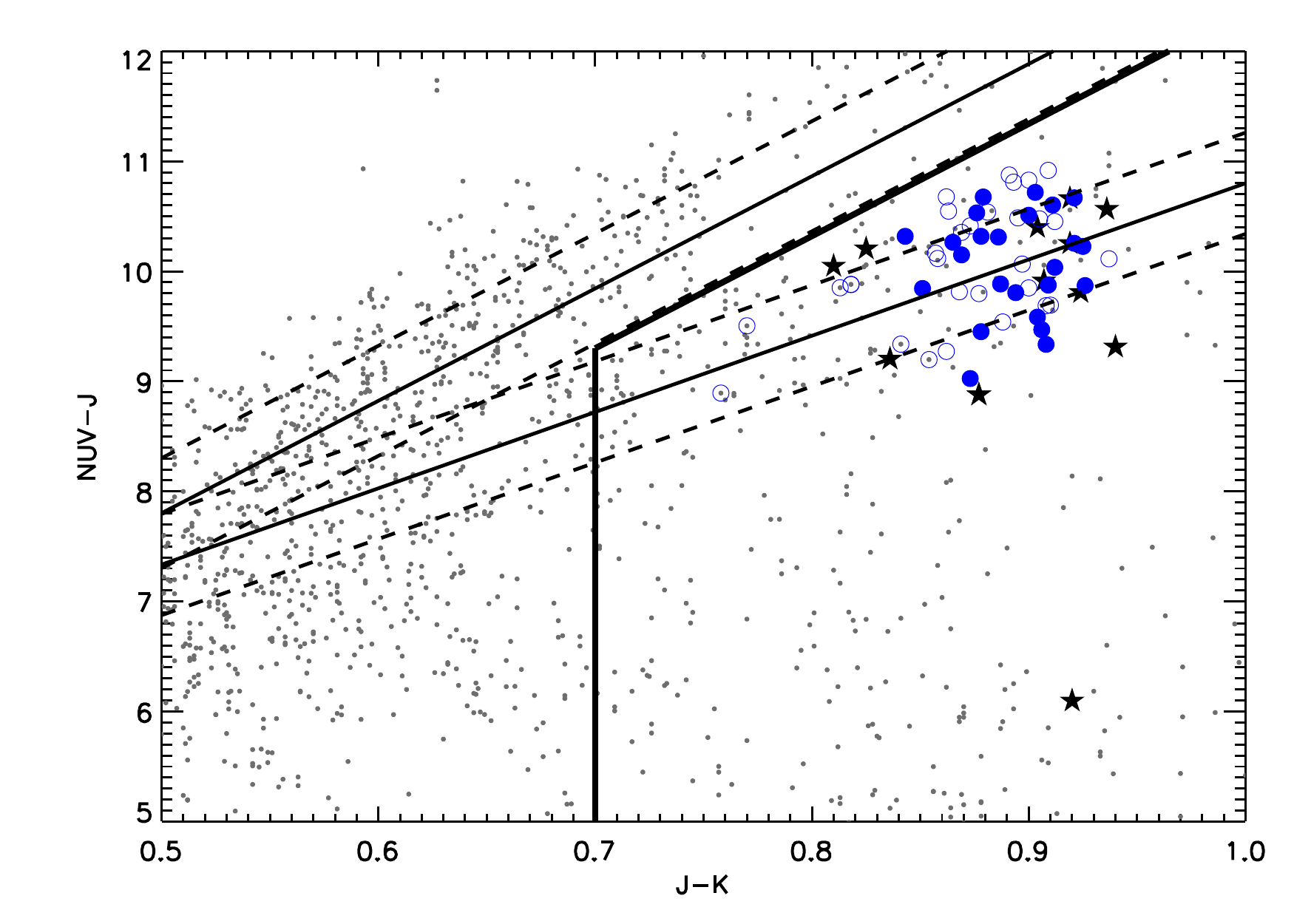}
\end{center}
\caption{
Our 54 UV-excess candidate stars are plotted as circles with 
filled circles denoting our spectroscopic sample.
TWA members known previous to our study (namely, those listed in \citealt{Torres:2008}) 
are plotted with star symbols.
Our selection criteria are denoted by the thick black lines; 
the thin solid lines denote the relation in \citet{FH10} 
for the stellar sequence and 
our best-fit line to the young stars (see Fig.~\ref{fig:ages}). 
Both have their associated 1-$\sigma$ errors as dashed lines.
The grey dots are the GALEX/2MASS sources in the vicinity of 
TWA with J$\leq$14, which removes many of the galaxies.
While we select all objects that satisfy our color-color criteria, most do not have 
proper motions, distances, or UVW consistent with 
those of young stars (see Sections~\ref{method:pm} and~\ref{results:id}).
}
\label{fig:selection}
\end{figure}

In Figures~\ref{fig:colors} and~\ref{fig:ages}, the young stars of the TWA
all lie below the stellar sequence.
However, there is some scatter, and not every 
object with UV excess will be a young star.
For example, flare stars are known contaminants 
in young star X-ray searches, and the same is true for our 
UV-selection methods (see also Figs.~\ref{fig:nuvabs} and~\ref{fig:nuvv}).
We select objects with UV excesses similar to those of the known young stars 
(defined algebraically by NUV--J $\leq$ 10.20(J--K) $+$ 2.2) 
and J--K between 0.7 and 1, as most M-type stars have 
J--K colors of about 0.8. Our selection criteria is illustrated in Figure~\ref{fig:selection} 
along with our candidate objects (see \S~\ref{results}) and fits to the old and young 
stellar locus as described above (see also Fig~\ref{fig:ages}).

\subsection{Proper Motions and UVW Space Velocities} \label{method:pm}

To distinguish between candidate young stars and other UV excess sources, we make use 
of proper motion information available in the literature.
In particular, we use the USNO-B1 \citep{Monet:2003}, 
UCAC3 \citep{Zacharias:2010}, and 
PPMX \citep{Roser:2008} proper motion catalogs, which can be accessed from the 
TOPCAT\footnote{Available at http://www.starlink.ac.uk/topcat/} Virtual Observatory tool.
In some cases we also check the PPMXL catalog \citep{Roeser:2010}.
Our UV excess sources are matched against all three catalogs, 
compared with proper motions of members in the association 
of interest, and then merged while removing duplicate entries.
In cases where one catalog has a proper motion that is discrepant from the others, 
we favor the proper motions listed in UCAC3 and PPMX.

We next estimate galactic space velocities (UVW) for candidate stars. 
This step requires estimates for distances and radial velocities. 
In order to obtain photometric distances, we estimate spectral types using 
available broadband photometry
(for example, data from DENIS \citep{DENIS} 
or the proper motion catalogs listed above).
From the spectral type we estimate photometric distances using empirical and theoretical 
isochrones for young stars of ages 10 and 100~Myr 
\citep[and references therein]{Song:2003,ZS04}.
Because nearly all candidates lack radial velocities, we use these distance estimates 
(along with the positions and proper motions) to estimate Galactic UVW 
with respect to the Sun for a range of velocities extending from $-80$ to 80~km/s.
\citet{ZS04} define a ``good UVW box" that contains nearly all young stars.
If any object has UVW within this box, 
that is, UVW within 0 to $-15$, $-10$ to $-34$, and $+3$ to $-20$~km~s$^{-1}$,
they are flagged for followup.
In this case U is defined as positive towards the Galactic center.

Follow up for all candidates so identified involves first 
checking against known association members.
In the TWA, for example, nearly all known members are 
recovered via the above methodology, including some 
that are not included in \citet{Torres:2008} but are present in older compilations, 
such as those in \citet{ZS04}.
The candidates are searched in SIMBAD as well, though most have no SIMBAD entry 
and only limited information in VizieR.

\subsection{Lithium Line Spectroscopy}

Candidates considered for spectroscopic followup are those 
with a broad range of radial velocities that would give UVW 
velocities consistent with nearby, young star status and 
photometric distances consistent with other members in the association of interest.
The presence and strength of Li 6708\AA~absorption 
in the spectrum can indicate whether or not 
the object is young \citep[see Figure~3 in][and 
the associated discussion in \S~5.1 of the same paper]{ZS04}.
A high-resolution spectrum can also be used to determine the radial velocity of 
the candidate object, which can then be used to better constrain its UVW velocities.

Although the presence of Li for stars of the spectral types of interest (K5 and later) 
is a strong indication of youth, 
lithium burning is very sensitive to stellar mass (and thus spectral type).
For example, $\beta$~Pic Moving Group stars ($\sim$12~Myr) of spectral types M3--4 
show very little Li in their spectra \citep{Song:2002,ZS04,Yee:2010}.
Hence, a non-detection of lithium absorption 
does not immediately imply the system is old, particularly for later M spectral types.
As demonstrated by \citet{Yee:2010}, lithium depletion models tend to predict 
much weaker Li absorption at any given age.
Using such models would yield far greater ages for $\sim$12~Myr M3.5--M5 stars.
Because many of our candidates fall in this range, a non-detection would imply 
a system is older than about 10~Myr, but does not provide an upper limit to this age.
Such a system could, however, still be a member of a young association.
In such cases, membership could be confirmed via  
accurate distance, proper motion, and radial velocity measurements.

\section{Application: Identification of Nearby Young Stars in and near the TW Hya Association}\label{results}

\subsection{NUV/NIR-based Candidate Identification} \label{results:id}

\begin{figure}[htb]
\begin{center}
\includegraphics[width=14cm,angle=0]{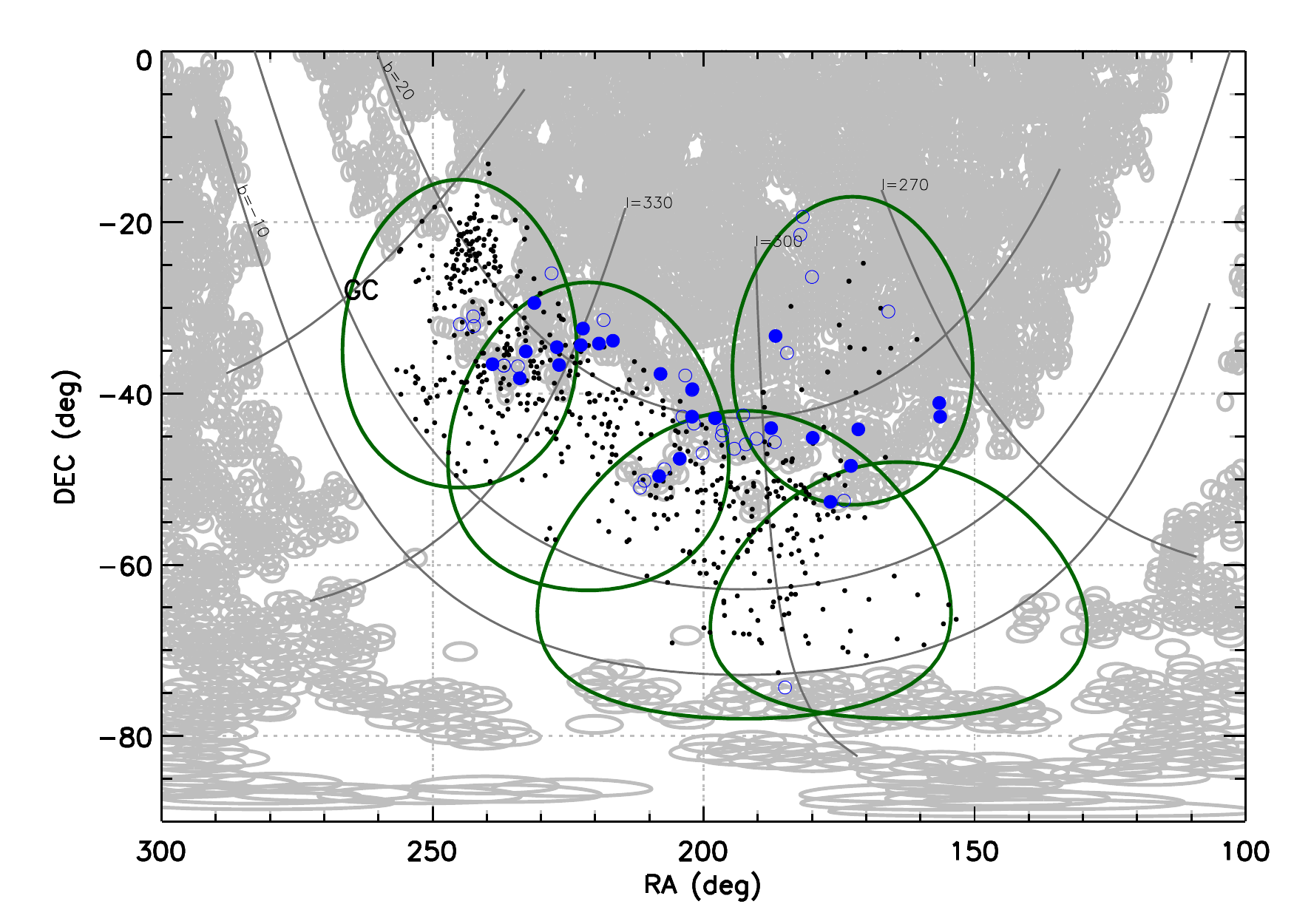}
\end{center}
\caption{Search regions (see Table~\ref{tab:fields}) for the Sco--Cen 
and TW Hya Associations.
Dots are candidate TWA, Upper Scorpius, UCL, and LCC members from \citet{Torres:2008} 
and \citet{deZeeuw:1999}.
Circles denote candidate objects (Table~\ref{tab:stars}); 
those we have spectra for are displayed as filled circles (Table~\ref{tab:ews}).
Our search regions encompasses all known members and candidates 
for these regions and extends somewhat beyond 
in order to explore for more distant members.
All GALEX fields in the GR4/5 database are denoted in gray.
Note that GALEX avoids the galactic plane (generally within 10$^\circ$) so that our coverage 
is better on the northern parts of our searched fields.
}
\label{fig:fields}
\end{figure}

We have used the method described in Section~\ref{method} to perform 
a search for young stars in the TW Hya Association 
(TWA, $\sim$8~Myr, 40--60~pc; \citealt{ZS04,Torres:2008}, and references therein) 
and the Scorpius--Centaurus region \citep{PM:2008}.
The Sco--Cen region is divided into a number of associations: 
Upper Scorpius (US, $\sim$5--6~Myr old), 
Upper Centaurus-Lupus (UCL, $\sim$14--15~Myr), 
and Lower Centaurus-Crux (LCC, $\sim$11--12~Myr), 
which are located about 120--140~pc distant 
\citep{PM:2008}\footnote{Based on Li abundances in late-type 
members of LCC and UCL, Song, Bessel, \& Zuckerman 
(unpublished) suggest these regions are only $\sim$10~Myr old.}.
In order to limit our searches to a manageable number of objects 
(the GALEX catalog CasJobs search interface has built-in memory limits), 
we divided the Sco--Cen search into a set of 4 RA/Dec centroids, 
each with a given radius (see Table~\ref{tab:fields} and Fig.~\ref{fig:fields}).
There is some overlap between these regions, so the resulting lists of objects 
(after cross-correlating with 2MASS) 
were merged and duplicate entries were removed.
Candidate objects were selected via the technique described in 
Sections~\ref{method:uv} and \ref{method:pm}.

We identified $\sim$150 promising candidates through this methodology. 
In order to select the best objects to observe, we selected only those with a broad range of 
possible radial velocities that would give UVW consistent with those of young stars, 
and those with distances comparable to that of the TWA or Sco--Cen regions.
After these cuts we had selected 10 TWA candidates and 44 Sco--Cen candidates.
These 54 candidates are listed in Table~\ref{tab:stars}, with spectral 
types estimated from photometric colors.
These candidate objects are also displayed in Figure~\ref{fig:selection} 
along with known TWA members for comparison.
Note, however, that the LCC and TWA are very close 
in the sky and have similar ages, 
so accurate distances are needed to 
determine in which association these targets belong.
The TWA and Sco--Cen search fields are displayed in Figure~\ref{fig:fields}, 
with candidate objects shown as open circles and those for which 
we obtained spectra (Table~\ref{tab:ews}) shown as filled circles.
 
\subsection{Spectroscopic Followup}\label{observations}

\begin{figure}[htb]
\begin{center}
\includegraphics[width=14cm,angle=0]{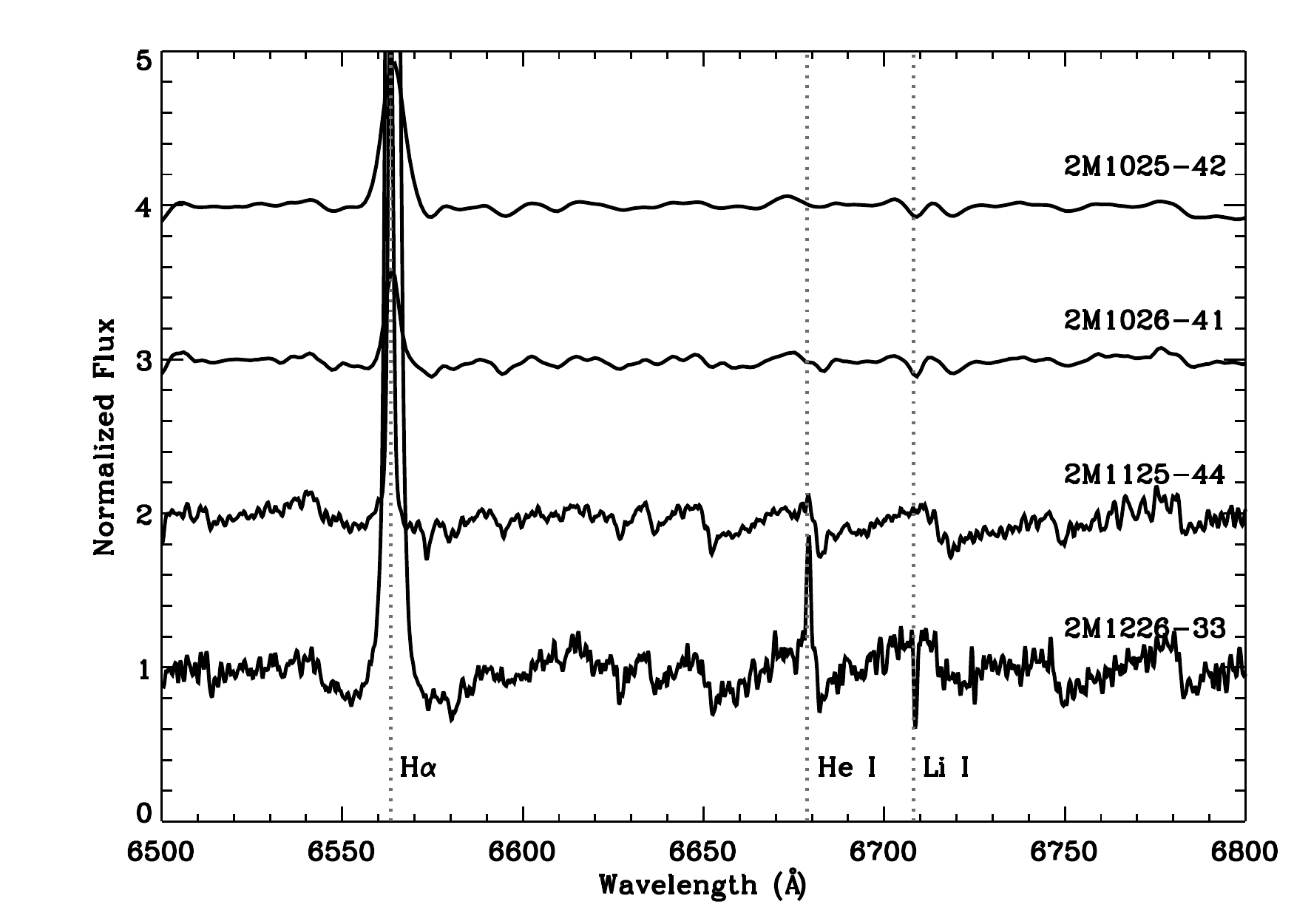}
\end{center}
\caption{$R_{3000}$ WiFeS spectra for our 4 TWA candidates.
The spectra for 2M1125-44 and 2M1226--33 were obtained with R=7000.
}
\label{fig:spectra}
\end{figure}

\begin{figure}[htb]
\begin{center}
\includegraphics[width=10cm,angle=0]{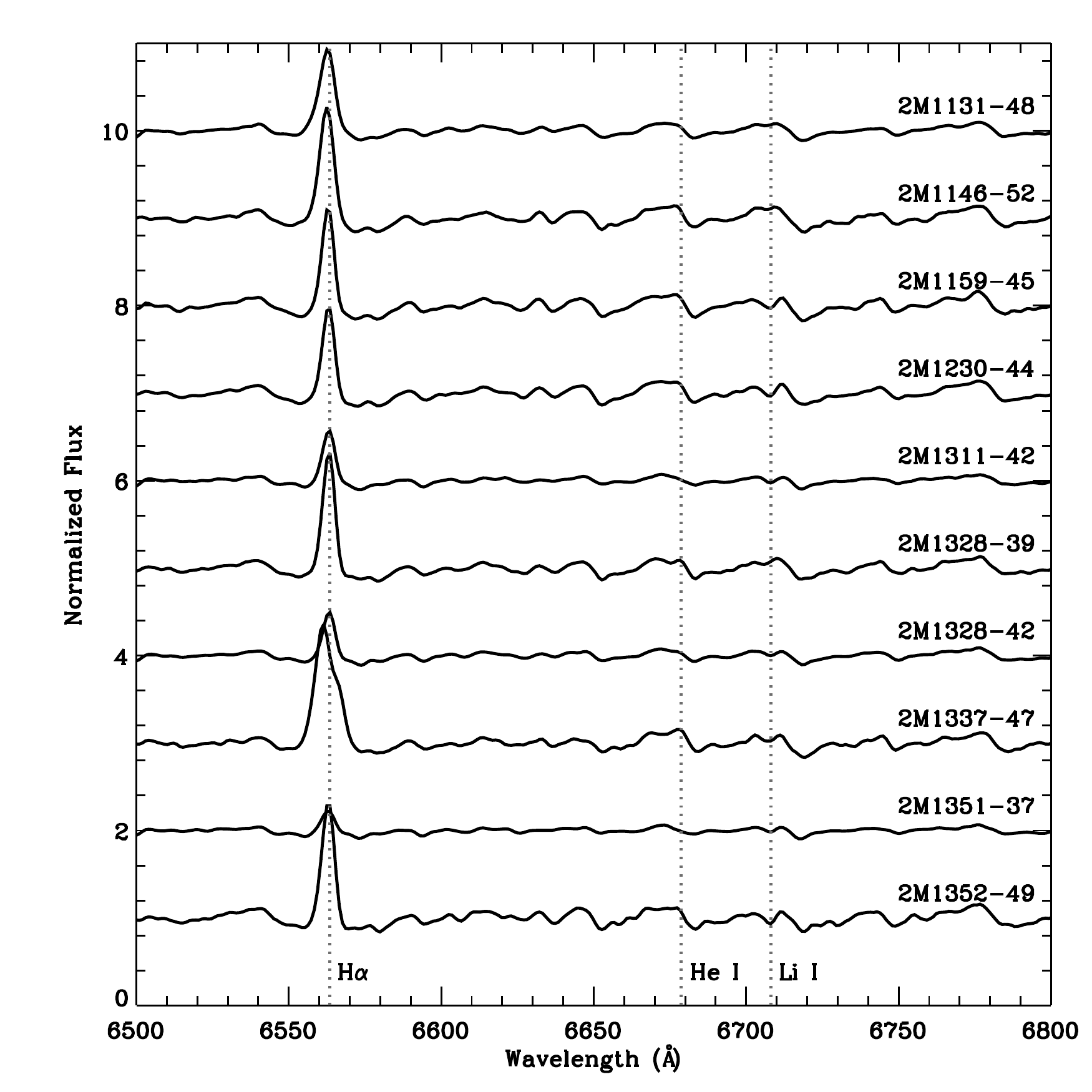}
\includegraphics[width=10cm,angle=0]{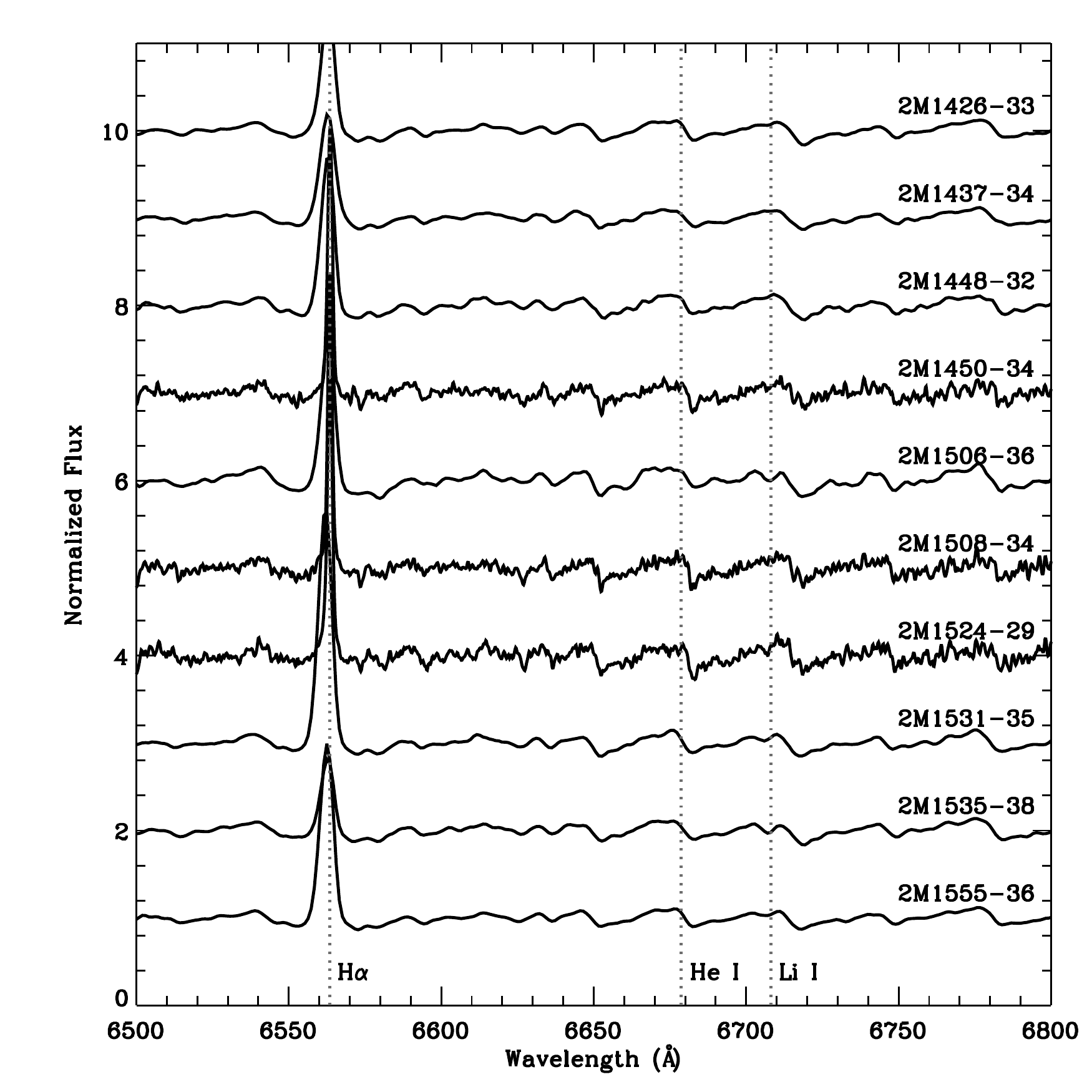}
\end{center}
\caption{WiFeS spectra for our 20 Sco--Cen candidates.
2M1337-47, which shows asymmetry in its (blue-shifted) H$\alpha$ profile, also shows a  
double peak for the other Hydrogen lines possibly due to 
accretion from a surrounding disk \citep{Kurosawa:2006}.
The spectra for 2M1450--34, 2M1508--34, and 2M1524--29 were obtained with R=7000.
}
\label{fig:spectra2}
\end{figure}

To check for signatures of youth, such as the presence of lithium 
absorption at 6708\AA, we obtained 
R=3000 spectra of the 24 stars listed in Table~\ref{tab:ews} with the 
Wide Field Spectrograph (WiFeS) at the 
Siding Springs Observatory's 2.3-m telescope during 2010 March and 2010 May. 
We also acquired R=7000 spectra for five of our objects in 2010 August.
Four of these targets were selected in our TWA 
search and 20 in our Sco--Cen searches, 
though a few Sco--Cen candidates overlap with 
the TWA search region (see Fig~\ref{fig:fields}).
WiFeS \citep{Dopita:2007} is a double--beam, 
image--slicing integral field spectrograph, and provides a $25\times38$ arcsec 
field with 0.5\arcsec~pixels.
A variety of gratings can be used to provide R=3000 and 
R=7000 spectra over optical wavelengths.
We used the $B_{3000}$ and $R_{3000}$ gratings to 
cover the wavelength range from 3200\AA~to 9800\AA.
The spectra for the region covering H$\alpha$ and Li 6708\AA~are displayed 
in Figures~\ref{fig:spectra} and~\ref{fig:spectra2}.
We have measured equivalent widths of H$\alpha$, He I, and Li I for these objects 
(when detected) and summarize these values in Table~\ref{tab:ews}.
Spectral types listed in Table~\ref{tab:ews} are estimated 
using the TiO5 index as described in \citet{Reid:1995}, 
with an estimated precision of $\pm0.5$ in subclass.

\subsection{Results and Discussion}\label{discussion}

Of the 24 targets listed in Table~\ref{tab:ews}, 
17 display evidence of Li absorption at 6708\AA.
The remaining 7 have no Li or an unidentified weak/moderate feature is present 
at a slightly bluer wavelength (offset by 2--3\AA).
Those objects with the weak feature near Li were re-observed with R=7000 (see below).
Note, however, that all systems with no evidence of Li in their spectra 
have spectral types M3.5--M4.5.
As previously mentioned, stars of these spectral types will burn lithium 
in as little as $\sim$10~Myr such that a lack of lithium 
does not immediately imply the system is old.
Of those that do display some lithium absorption, 6 are M1--M3, 
with the remaining being later than M3.5.
Those later than M3.5 must be at most $\sim$10~Myr old in order 
to show such strong lithium absorption.
The earlier M-types will burn their lithium somewhat 
more slowly, so these systems may be of age $\sim$10~Myr.

X-ray detection has been one way young stars have 
been identified \citep[for example,][]{Kastner:1997,Song:2003,SACY}.
Only 5 of our 24 spectroscopic targets (and only 14 among the 54 candidates) 
have ROSAT All-Sky Survey (RASS) X-ray detections within 1\arcmin~of 
the target's 2MASS coordinates in 
either the Bright or Faint source catalogs. 
The lack of X-ray detection in RASS suggest these objects, if they are young, 
are producing too little X-ray emission to be readily detected at these distances.
Table~\ref{tab:stars} lists $L_X/L_{bol}$ for all candidates, including upper limits 
for RASS non-detections (assuming a RASS flux limit of 
$2\times10^{-13}$~ergs~cm$^{-2}$~s$^{-1}$; \citealt{Schmitt:1995}).
All RASS-detected candidates have log~$L_X/L_{bol} \sim -3$, which is typical of 
nearby M-stars (see Fig.~5 in \citealt{Riaz:2006}).

H$\alpha$ equivalent widths (EW) is one way that T Tauri stars are identified and classified.
For these spectral types, values for H$\alpha$ EW $\geq$10--20\AA~are typically
observed in classical T Tauri stars \citep{White:2003}.
Of the stars observed spectroscopically, 9 show H$\alpha$ EW larger than 10\AA, 
with a 10th system having an H$\alpha$ EW consistent with 10\AA.
In addition to H$\alpha$, we also measure He~I EW at 5876\AA~and 6678\AA~in 
those systems with clear He~I emission. 
The 8 systems with measured He~I also have strong H$\alpha$.
Of these, 3 have no lithium absorption detected, but have spectral types M3.5--M4. 
The detected He~I and lack of Li may suggest these systems are young, 
but somewhat older than $\sim$10~Myr.
We comment on several noteworthy targets below for 
which we acquired R=7000 spectra.

{\bf 2M1125--44}: This is an M4 star detected in our TWA search, 
with a photometric distance of 75~pc if 10~Myr-old.
The R=3000 spectrum suggested a dip near 6708\AA, 
but a higher resolution spectrum revealed no detectable Li, 
though He I emission lines at 5876 and 6678\AA~are detected. 
The radial velocity for this system 
($19.5 \pm 2$ km s$^{-1}$; S. Murphy, private comm.), 
combined with the 10~Myr-old photometric distance estimate, 
provide UVW that are consistent with those of young stars.
At a distance of 75~pc, the Galactic UVW are $-4\pm5$, $-23\pm3$, 
$-4\pm5$~km~s$^{-1}$.
The lack of Li, however, implies that the star is older than $\sim$10~Myr, but even 
at $\sim$30~Myr (implying a closer distance of about 50~pc)
the UVW are still consistent with those of a young star 
($-1\pm4$, $-22\pm2$, $-1\pm4$~km~s$^{-1}$).
This star has been detected in X-rays by the ROSAT All Sky Survey 
and has log $L_X/L_{bol}\sim-2.33$.
Hence, this candidate is possibly a young star 
based on dynamics and X-ray emission, 
though is not likely to be a member of the $\sim$8~Myr TWA.
The UVW is marginally consistent with that of TWA ($-11$, $-18$, $-5$~km~s$^{-1}$; \citealt{ZS04}) and the Sco-Cen region (UCL: $-6.8$, $-19.3$, $-5.7$, LCC: $-8.2$, $-18.6$, $-6.4$, with dispersions of 3--7~km~s$^{-1}$; \citealt{Sartori:2003}).
A parallax measurement will be key in determining the age of this object and 
membership in any nearby association.

{\bf 2M1226--33}: This is an M5 star detected in our TWA search 
with a photometric distance of 65~pc if 10~Myr-old.
This system has very strong H$\alpha$ and He lines.
The H$\alpha$ has equivalent width of $64 \pm 4$\AA~and
the 5876\AA~and 6678\AA~He I lines have EWs $\sim$6\AA~and $\sim$1\AA, respectively.
Strong lithium absorption is also observed (410~m\AA).
The presence of such strong lines suggests the system 
may be a classical T~Tauri star \citep{White:2003}.
We obtained an additional spectrum for this star using R=7000.
The equivalent widths differ somewhat from the R=3000 spectrum and are listed 
as an additional row in Table~\ref{tab:ews}.
\citet{White:2003} find that H$\alpha$ full widths at 10\% higher 
than 270~km~s$^{-1}$ are found among classical T~Tauri stars.
With the higher resolution spectrum, we measured a full width at 
the 10\% level for H$\alpha$ of 300~km~s$^{-1}$, suggesting 
this object could be a nearby classical T~Tauri system.
Comparison with Figure~3 in \citet{Natta:2004} suggests 
an accretion rate of $\sim10^{-10}$~$M_\odot$~$yr^{-1}$.
With the higher resolution spectrum, we were also able to 
determine a radial velocity of $14.8\pm3$ km s$^{-1}$.
We calculate UVW of $-8\pm3$, $-24\pm4$, $-1\pm3$~km~s$^{-1}$.
The UVW is marginally consistent with that of TWA and Sco-Cen.
There is no RASS detection for this object implying log~$L_{X}/L_{bol}<-3.10$; 
the field has not been observed with Chandra or XMM. 

{\bf 2M1450--34, 2M1508--34, \& 2M1524--29}: The R=3000 spectra 
for these three stars had an absorption feature near lithium, 
but offset by 2--3\AA~from the expected 6708\AA.
A higher resolution spectrum reveals no clear signature of lithium, 
though a blended $\sim$100~m\AA~feature is seen in 
2M1524--29 at a wavelength of 6707.1\AA~(see Figure~\ref{fig:spectra2}).
2M1450--34's R=3000 spectrum shows possible He I emission at 6678\AA, 
but this is not seen in the R=7000 spectrum
(the R=7000 spectrum does not cover the region where He I 5876\AA~can 
be seen in emission).
The radial velocities measured from the R=7000 spectrum are 
7.3, 4.9, and 6.0~km~s$^{-1}$, (all with uncertainties of $\pm3$~km~s$^{-1}$) 
for 2M1450--34, 2M1508--34, and 2M1524--29, respectively.
With 10~Myr photometric distance estimates, the UVW for 2M1450--34 
are: $-5\pm3$, $-29\pm6$, $-6\pm3$~km~s$^{-1}$;
for 2M1508--34: $-7\pm3$, $-29\pm6$, $-5\pm3$~km~s$^{-1}$;
and for 2M1524--29: $-1\pm3$, $-23\pm5$, $-3\pm3$~km~s$^{-1}$.
Both 2M1450--34 and 2M1508--34 have UVW that appear to be marginally consistent with those of TWA and Sco-Cen.
None of these objects have detections in RASS 
(note that all three stars have photometric distances larger than 100~pc, 
see Table~\ref{tab:stars}), nor 
have they been observed by Chandra or XMM.

\subsection{Two Stars with Unusual NUV--J Color}\label{additional}

While carrying out the search described in Sections~3.1--3.3, 
we found two systems in \citet{Torres:2008} 
with particularly unusual NUV--V and NUV--J colors (Figs.~\ref{fig:nuvv} and~\ref{fig:ages}):

{\bf BD +01 2447}: This M2 star is listed in \citet{Torres:2008} as an AB~Dor 
Association member (age 70~Myr).
BD~+01 2447, also known as HIP~51317, is 7~pc away,
with an upper limit of 60~m\AA~for Li EW \citep{Zickgraf:2005}.
For an M2 star, however, lithium may be depleted within $\sim$12~Myr 
(see Fig.~3 in \citealt{ZS04}).
The kinematics of the system are consistent with membership in 
the AB~Dor Association and X-ray emission has been detected 
(log $f_X/f_V \sim -3.53$; \citealt{Zickgraf:2005}).
This object has a 2MASS K magnitude of 5.311, 
which (at 7~pc) corresponds to $M_K$ of 6.1.
The models of \citet{Baraffe:1998} predict an M2 star 
would reach such an $M_K$ in about 80--100~Myr.
These results are 
consistent with this being a nearby AB~Dor member.
However, in Figures~\ref{fig:nuvabs}, \ref{fig:nuvv}, and \ref{fig:ages}, 
this system is located in the region of color-color (and color-magnitude) space 
more common for older systems.
It is noteworthy that, of all of the $\leq$70~Myr stars in \citet{Torres:2008} 
detected by GALEX, BD~+01 2447 is the only star (excepting TW~Hya, see below) 
that lies significantly outside the locus of young stars.
BD +01 2447 also displays slow rotation ($v$~sin$i<3$~km~s$^{-1}$), which is more commonly seen in older systems (T.\ Forveille 2010, private communication).
A closer examination of BD~+01~2447 is warranted so as 
to understand its weak NUV emission.

{\bf TW Hya}: This star, namesake of the TW Hya Association, 
stands out in all our NUV figures due to its substantial NUV excess.
In particular, it has NUV--J$\sim$6, while the rest of the TWA stars 
have NUV--J$\sim$9--11.
If the UV excesses displayed by the TWA stars can be ascribed to 
chromospheric activity due to their youth, 
an additional emission source is required to explain 
why TW Hya is $\sim$3 magnitudes brighter in NUV than the rest of the TWA.
Of the young stars plotted on our figures, 
TW Hya is the only known classical T~Tauri star.
The most likely explanation for the increased UV emission 
is ongoing accretion onto the star, as suggested 
by its optical and X-ray emission properties 
($\sim10^{-8}$~$M_\odot$~$yr^{-1}$, \citealt{Kastner:2002}).
Additionally, the orientation of the system is such that we are seeing its accretion 
unobscured (the disk is near face-on, \citealt{Krist:2000}).
Our unobscured view of the accretion onto this star is therefore the likely 
explanation for its high UV flux.
This suggests that, barring intervening absorbing material, 
actively accreting stars such as TW~Hya will 
readily stand out in GALEX-Optical/NIR colors.

\section{Conclusions}\label{conclusions}

We have demonstrated a new technique based on GALEX and 2MASS colors 
that can readily identify young, low-mass stars near the Earth.
\citet{FH10} similarly demonstrated that such a UV/near-IR search strategy 
is feasible for stars younger than $\sim$5~Myr, 
but essentially no such stars are known to exist 
within $\sim$100~pc of Earth.
The present study demonstrates that GALEX NUV fluxes 
may be used in conjunction with either optical or near-IR fluxes 
to identify nearby stars with ages in the range 10--100~Myr 
and spectral types K5 or later.
With additional information, namely space velocities, 
we have formulated an efficient method to identify
candidate members of nearby stellar associations.

We have performed a search in the TWA and Sco--Cen region and 
recovered 54 high-quality candidates.
Of these, we observed 24 spectroscopically and 
detected lithium absorption, an indicator of youth, in at least 17 M1--M5 systems.
Those with no lithium absorption have spectral types M3.5--M4.5, 
a range where lithium is believed to be depleted 
within $\sim$10~Myr \citep{Song:2002,ZS04,Yee:2010}.
These systems may very well be young, albeit not younger than 10~Myr.
Detection of He~I emission at 5876\AA~and 6678\AA~for 8 systems 
(some of which had no lithium absorption) is also consistent with 
young ages for these stars.

The detection of lithium in at least 2/3 of the 24 spectroscopically observed stars 
suggests many of the other high-quality candidates may also 
be young stars, and perhaps others lie among the 100+ UV-excess stars 
with possible good UVW from which those candidates were drawn.
Most of our objects lack RASS detections within 1\arcmin~of their 2MASS coordinates; 
however, these tend to have upper limits of log~$L_X/L_{bol} \sim -3$, 
consistent with late-type stars that have been detected in the RASS.
Because mid to late M-type stars have small $L_{bol}$, 
many would escape detection in RASS even if log~$L_X/L_{bol} \sim -3$.
While deeper X-ray observations can be performed with Chandra 
or XMM, these will eventually cover only a few percent of the sky.
This suggests that UV-selection criteria, such 
as the one we describe in this paper, 
may ultimately prove a more powerful tool than X-rays 
to identify nearby, young, low-mass stars.

\acknowledgements
{\it Acknowledgements.} 
We thank Simon Murphy for providing the high resolution spectrum for 2M1125--44 
and the referee for useful suggestions.
This publication makes use of data products from GALEX, operated for NASA by the California Institute of Technology, and the Two Micron All Sky Survey, which is a joint project of the University of Massachusetts and the Infrared Processing and Analysis Center/California Institute of Technology, funded by the National Aeronautics and Space Administration and the National Science Foundation.
This work has used the SIMBAD and VizieR databases, operated at CDS, Strasbourg, France.
This research was supported by NASA Astrophysics Data Analysis Program grant NNX09AC96G to RIT and UCLA.


\begin{deluxetable}{lccccc}
\tablecaption{GALEX Searches}
\tablewidth{0pc}
\tabletypesize{\small}
\tablehead{
\colhead{Region\tablenotemark{a}} & \colhead{RA} & \colhead{Dec} & \colhead{Radius} & \colhead{GALEX/2MASS}\\
\colhead{} & \colhead{} & \colhead{} & \colhead{} & \colhead{sources}
}
\startdata
TWA   & 11h30m & $-35^\circ$ & $18^\circ$ & 47,101\tablenotemark{b}\\
LCC 1 & 10h56m & $-63^\circ$ & $15^\circ$ & 381,385 \\ 
LCC 2 & 12h50m & $-60^\circ$ & $18^\circ$ &  \\
UCL 1 & 14h45m & $-45^\circ$ & $18^\circ$ & 548,324 \\
UCL 2 & 16h20m & $-33^\circ$ & $18^\circ$ &  
\enddata
\tablenotetext{a}{TWA: TW Hydra Association; LCC: Lower Centaurus-Crux; UCL: Upper Centaurus-Lupus}
\tablenotetext{b}{Only the TWA search required detection in both NUV and FUV.}
\tablecomments{
\ List of all searches performed. See also Fig.~\ref{fig:fields}.
Number of stars quoted for LCC~1 region also includes stars in LCC~2; 
similarly for UCL~1 and UCL~2.
}
\label{tab:fields}
\end{deluxetable}

\begin{deluxetable}{lllcclrcccc}
\tabletypesize{\footnotesize}
\rotate
\tablecolumns{11}
\tablewidth{0pc}
\tablecaption{UV Selected Objects} 

\tablehead{
\colhead{RA} & \colhead{Dec} & \colhead{2MASS} & \colhead{$\mu_\alpha$} 
& \colhead{$\mu_\delta$} & \colhead{J--K} & \colhead{NUV-J} & \colhead{Spectral} 
& \colhead{D} & \colhead{RV Range} & \colhead{log $L_X/L_{bol}$} \\
\colhead{(deg)} & \colhead{(deg)} & \colhead{Designation} & \colhead{(mas/yr)} & 
\colhead{(mas/yr)} & \colhead{} & \colhead{} & \colhead{Type} & \colhead{(pc)} & 
\colhead{(km/s)} & \colhead{}}

\startdata
  156.337 & -42.6983 & 10252092--4241539 & -47. & -2. & 0.91 & 10.04 & M2-M2.5 & 91. & 10:30 & $-3.03$\\
  156.509 & -41.0983 & 10260210--4105537 & -46. & -3. & 0.9 & 9.58 & M3-M3.5 & 67. & 8:31 & $-3.24$\\
  165.8988 & -30.4137 & 11033571--3024494 & -36. & -18. & 0.84 & 9.34 & M0 & 126. & 0:26 & $<-3.57$\\
  171.448 & -44.1741 & 11254754--4410267 & -37. & -18. & 0.86 & 10.26 & M4-M4.5 & 75. & 2:27 & $-2.33$\\ 
  172.812 & -48.4411 & 11311482--4826279 & -40. & -6. & 0.87 & 9.03 & M3-M3.5 & 130. & 17:26 & $<-3.24$\\
  174.079424 & -52.515057 & 11361906--5230542 & -31. & -10. & 0.86 & 10.12 & M3-M3.5 & 123. & 4:29 & $-2.80$\\
  176.596286 & -52.647766 & 11462310--5238519 & -46. & -14. & 0.89 & 9.88 & M4.5-M5 & 85. & 3:28 & $<-2.97$\\
  179.866105 & -45.172005 & 11592786--4510192 & -44. & -24. & 0.87 & 10.15 & M4.5-M5 & 55. & 5:21 & $-2.99$\\
  179.99 & -26.3761 & 11595770--2622340 & -32. & -17. & 0.85 & 9.20 & M3-M3.5 & 113. & -2:23 & $<-3.35$\\
  181.698 & -19.3481 & 12064743--1920531\tablenotemark{a} & -56. & -7. & 0.77 & 9.50 & K7-M0 & 71. & 2:11 & $<-3.57$\\
  182.1542 & -21.45811 & 12083700--2127291\tablenotemark{a} & -73. & -36. & 0.81 & 9.85 & M3 & 60. & -4:21& $<-2.92$\\
  184.598 & -35.2527 & 12182363--3515098 & -28. & -13. & 0.82 & 9.88 & M0.5-M1 & 101. & 2:22 & $-3.19$\\
  184.973163 & -74.33593 & 12195355--7420093 & -34. & -14. & 0.89 & 10.81 & M3.5-M4 & 104. & 5:28 & $-3.03$\\
  186.714 & -33.2701 & 12265135--3316124 & -62. & -25. & 0.91 & 9.34 & M5 & 65. & -2:23 & $<-3.10$\\
  186.855 & -45.6685 & 12272529--4540065 & -30. & -14. & 0.76 & 8.89 & K7-M0 & 149. & 2:25 & $-3.11$\\
  187.521722 & -44.04332 & 12300521--4402359 & -44. & -18. & 0.88 & 10.53 & M4.5-M5 & 69. & 2:24 & $<-3.25$\\
  190.225488 & -45.27367 & 12405411--4516252 & -35. & -13. & 0.91 & 10.92 & M3-M3.5 & 124. & 2:26 & $<-3.19$\\
  192.215415 & -45.939037 & 12485169--4556205 & -27. & -14. & 0.9 & 10.07 & M4.5-M5 & 121. & 0:23 & $<-2.76$\\
  192.698969 & -42.53022 & 12504775--4231487 & -34. & -18. & 0.82 & 9.88 & M4.5-M5 & 120. & -1:26 & $<-2.76$\\
  194.322195 & -46.447998 & 12571732--4626527 & -31. & -20. & 0.87 & 10.35 & M3.5-M4 & 146. & 2:23 & $<-2.95$\\
  196.442675 & -44.293636 & 13054624--4417370 & -35. & -16. & 0.89 & 10.49 & M3-M3.5 & 143. & 6:22 & $-2.78$\\
  196.632647 & -44.919258 & 13063183--4455093 & -29. & -13. & 0.87 & 10.41 & M2-M2.5 & 138. & 0:26 & $<-3.28$\\
  197.870938 & -42.878288 & 13112902--4252418 & -35. & -19. & 0.9 & 10.51 & M2-M2.5 & 121. & 0:25 & $-2.93$\\
   \multicolumn{11}{c}{Continued on next page}
   \tablebreak
  200.17506 & -46.976494 & 13204201--4658353 & -29. & -20. & 0.91 & 9.69 & M4-M4.5 & 126. & -3:21 & $<-2.92$\\
  201.775952 & -43.506329 & 13270622--4330227 & -35. & -17. & 0.86 & 10.16 & M4.5-M5 & 118. & 0:24 & $<-2.80$\\
  202.060031 & -39.487438 & 13281440--3929147 & -28. & -32. & 0.91 & 10.60 & M4-M4.5 & 84. & -2:11 & $<-3.22$\\
  202.126198 & -39.568535 & 13283028--3934067 & -32. & -14. & 0.91 & 9.69 & M4-M4.5 & 134. & 0:21 & $<-2.86$\\
  202.135325 & -42.695457 & 13283247--4241436 & -44. & -36. & 0.93 & 10.23 & M3-M3.5 & 111. & 3:18 & $<-3.27$\\
  203.373612 & -37.88876 & 13332966--3753195 & -23. & -20. & 0.91 & 10.45 & M4-M4.5 & 143. & -7:16 & $<-2.89$\\
  203.880003 & -42.696892 & 13353120--4241488 & -26. & -23. & 0.94 & 10.11 & M4-M4.5 & 123. & -5:17 & $<-3.02$\\
  204.409991 & -47.608257 & 13373839--4736297 & -32. & -23. & 0.91 & 9.47 & M4.5-M5 & 126. & 0:23 & $<-2.70$\\
  207.22411 & -48.84594 & 13485378--4850453 & -33. & -18. & 0.86 & 10.68 & M3.5-M4 & 146. & 4:22 & $<-3.02$\\
  207.940297 & -37.700115 & 13514567--3742004 & -26. & -36. & 0.92 & 10.25 & M2-M2.5 & 119. & -9:13 & $<-3.43$\\
  208.190309 & -49.642429 & 13524567--4938327 & -37. & -17. & 0.84 & 10.32 & M4.5-M5 & 112. & 1:22 & $<-2.75$\\
  210.906607 & -50.179924 & 14033758--5010477 & -33. & -18. & 0.86 & 10.55 & M4-M4.5 & 98. & -3:17 & $<-3.16$\\
  211.712131 & -50.98185 & 14065091--5058546 & -28. & -24. & 0.9 & 10.83 & M4-M4.5 & 121. & -3:18 & $<-3.04$\\
  216.715646 & -33.826576 & 14265175--3349356 & -28. & -32. & 0.9 & 10.72 & M3.5-M4 & 139. & -5:14 & $<-3.05$\\
  218.440042 & -31.41147 & 14334561--3124412 & -20. & -30. & 0.89 & 9.54 & M4.5-M5 & 97. & -8:5 & $<-2.88$\\
  219.305101 & -34.155811 & 14371322--3409209 & -32. & -31. & 0.89 & 9.81 & M3-M3.5 & 145. & -3:15 & $<-3.17$\\
  222.245008 & -32.402317 & 14485880--3224083 & -34. & -32. & 0.85 & 9.84 & M4.5-M5 & 132. & -4:14 & $-2.38$\\
  222.639972 & -34.341827 & 14503359--3420305 & -29. & -31. & 0.88 & 9.45 & M4-M4.5 & 142. & -11:7 & $<-2.90$\\ 
  226.646079 & -36.658264 & 15063505--3639297 & -34. & -25. & 0.93 & 9.87 & M5-M5.5 & 118. & -4:13 & $<-2.58$\\
  227.068891 & -34.579636 & 15081653--3434466 & -31. & -32. & 0.91 & 9.87 & M4-M4.5 & 140. & -5:12 & $<-2.86$\\ 
  228.050838 & -25.952211 & 15121220--2557079 & -20. & -33. & 0.89 & 10.88 & M2-M2.5 & 99. & -14:3 & $<-3.58$\\
  231.234692 & -29.413294 & 15245632--2924478 & -26. & -32. & 0.88 & 10.32 & M4.5-M5 & 116. & -10:6 & $<-2.76$\\
  232.789928 & -35.082542 & 15310958--3504571 & -34. & -30. & 0.92 & 10.67 & M4.5-M5 & 70. & -10:6 & $<-3.12$\\
  \multicolumn{11}{c}{Continued on next page}
  \tablebreak
  233.912086 & -38.210953 & 15353890--3812394 & -30. & -38. & 0.89 & 10.31 & M4-M4.5 & 126. & -5:11 & $<-2.94$\\
  234.257429 & -36.814156 & 15370178--3648509 & -16. & -34. & 0.88 & 9.80 & M4-M4.5 & 108. & -11:5 & $<-3.14$\\
  236.847334 & -36.736374 & 15472336--3644109 & -26. & -29. & 0.87 & 9.81 & M4.5-M5 & 135. & -7:9 & $<-2.68$\\
  236.852005 & -36.726772 & 15472448--3643363 & -24. & -30. & 0.86 & 9.27 & M3-M3.5 & 130. & -8:8 & $-3.00$\\
  238.948557 & -36.566711 & 15554765--3634001 & -30. & -37. & 0.88 & 10.68 & M4-M4.5 & 103. & -8:7 & $-2.86$\\
  242.36897 & -32.104786 & 16092855--3206172 & -12. & -34. & 0.9 & 9.85 & M4.5-M5 & 115. & -13:2 & $<-2.79$\\
  242.478602 & -30.982832 & 16095486--3058581 & -18. & -32. & 0.91 & 10.48 & M2-M2.5 & 130. & -11:4 & $<-3.36$\\
  244.974809 & -31.901608 & 16195395--3154057 & -24. & -26. & 0.88 & 10.54 & M3-M3.5 & 144. & -9:5 & $-2.86$\\
\enddata

\tablenotetext{a}{For these objects distance and RV ranges are quoted using 100-Myr isochrones.}
\tablecomments{
Spectral types are obtained from photometric colors.
RV Range is the range of radial velocities that give good UVWs (see Section~\ref{method:pm} for more details).
Unless otherwise noted in footnote $a$, all distances are calculated using 10-Myr isochrones and have uncertainties of 20\%.
Upper limits for the last column are derived assuming 
$F_X=2\times10^{-13}$ ergs cm$^{-2}$ s$^{-1}$, 
the characteristic RASS flux limit \citep{Schmitt:1995}.
}
\label{tab:stars}
\end{deluxetable}

\begin{deluxetable}{clrrrrrrr}
\tabletypesize{\footnotesize}
\tablecaption{Spectroscopic Results}

\tablehead{
  \colhead{Target} & \colhead{Spectral} & \colhead{Dist.} & \colhead{He I} 
  & \colhead{H$\alpha$} & 
  \colhead{He I} & \colhead{Li I}  \\
  \colhead{} & \colhead{Type} & \colhead{(pc)} 
  & \colhead{5876\AA} & \colhead{6563\AA} & 
  \colhead{6678\AA} &  \colhead{6708\AA} 
}

\startdata
10252092--4241539 & M1 & 91 &  & $-3.6\pm0.6$ &  & $490\pm30$ \\
10260210--4105537 & M1 & 67 &  & $-7.4\pm0.3$ &  & $500\pm70$ \\
11254754--4410267 & M4 & 75 & $-0.8\pm0.2$ &$ -8.2\pm0.2$ & $-50\pm30$ & $<30$ \\
12265135--3316124 & M5 & 65 & $-6.4\pm0.2$ & $-64.0\pm4.0$ & $-1200\pm200$ & $410\pm40$ \\
 				 &  &       &                       & $-51.1\pm7.0$ & $-670\pm110$  & $573\pm50$ \\
\hline
11311482--4826279 & M3 & 130 &  & $-7.9\pm0.5$ &  & $60\pm20$ \\ 
11462310--5238519 & M4.5 & 85 &  & $-10.1\pm0.5$ &  & $66\pm20$ \\ 
11592786--4510192\tablenotemark{a} & M4.5 & 55 &  & $-8.18\pm0.3$ &  & $527\pm50$ \\
12300521--4402359 & M4 & 69 &  & $-7.2\pm0.2$ &  & $503\pm30$ \\
13112902--4252418 & M1.5 & 121 &  & $-4.2\pm0.2$ &  & $255\pm20$ \\
13281440--3929147 & M4 & 84 &  & $-8.7\pm0.2$ &  & $165\pm20$ \\ 
13283247--4241436 & M2 & 111 &  & $-3.8\pm0.2$ &  & $180\pm20$ \\
13373839--4736297 & M3.5 & 126 & $-1.3\pm0.1$ & $-13.7\pm0.2$ & $-115\pm50$ & $308\pm40$ \\
13514567--3742004 & M1 & 119 &  & $-2.2\pm0.1$ &  & $200\pm10$ \\
13524567--4938327 & M4.5 & 112 &  & $-8.5\pm0.1$ &  & $517\pm50$ \\
14265175--3349356 & M3.5 & 139 & $-0.6\pm0.1$ & $-10.2\pm0.4$ & $-24\pm10$ & $<50$ \\
14371322--3409209 & M3.5 & 145 &  & $-9.7\pm0.6$ &  & $<30$ \\
14485880--3224083 & M4.5 & 132 &  & $-11.8\pm0.4$ &  & $<60$ \\
14503359--3420305 & M3.5 & 142 & $-1.0\pm0.1$ & $-10.5\pm0.4$ & $<-30$ & $<80$ \\
                                  &       &          &             & $-6.9\pm0.6$ & & $<50$ \\
15063505--3639297 & M5 & 118 & $-1.9\pm0.3$ & $-12.1\pm0.3$ & $-44\pm20$ & $446\pm90$ \\
15081653--3434466 & M3.5 & 140 &  & $-7.8\pm0.5$ &  & $<60$  \\
15245632--2924478 & M4.5 & 116 &  & $-8.7\pm0.6$ &  & $<100$ \\
15310958--3504571 & M4.5 & 70 & $-2.0\pm0.2$ & $-17.4\pm0.3$ & $-145\pm60$ & $258\pm40$ \\
15353890--3812394 & M4 & 126 &  & $-5.97\pm0.3$ &  & $388\pm40$ \\
15554765--3634001 & M3.5 & 103 & $-1.0\pm0.2$ & $-13.6\pm0.2$ & $-39\pm10$ & $184\pm20$ \\ 
\enddata

\tablenotetext{a}{The strong lithium line detected in 2M1159--45 was initially and
independently seen in previous observations at 
Siding Spring Observatory (Zuckerman et al.\ 2010, submitted to ApJ).}

\tablecomments{
The first four objects are from the TWA search, 
the rest are from the Sco--Cen searches (see Table~\ref{tab:fields}). 
Some candidate objects found in the Sco--Cen searches may be TWA candidates (see Fig.~\ref{fig:fields}).
Spectral types are estimated using the TiO5 index as described in \citet{Reid:1995} 
and distances assume the star is $\sim$10~Myr old.
He~I~(5876\AA) and H$\alpha$ equivalent widths are measured in \AA, all others are in milli\AA, with negative equivalent widths denoting emission lines. 
Upper limits (3-$\sigma$) are provided when a line is not detected.
The extra row below 2M1226--33 and 2M1450--34 are values measured from the R7000 spectra.
}
\label{tab:ews}
\end{deluxetable}

\clearpage

\end{document}